\documentclass[12pt, letterpaper]{article}

\usepackage[utf8]{inputenc}
\usepackage[T1]{fontenc}
\usepackage{setspace}
\usepackage{geometry}
\usepackage{natbib}
\usepackage{hyperref}
\usepackage{xcolor}
\usepackage{csquotes}

\title{\textbf{Consciousness, AI, and the Limits of Scientific Explanation}}

\author{
    Bradley C. Love\\
    \small EmpiriQal.ai\\
    \small \href{mailto:bradley.c.love@gmail.com}{bradley.c.love@gmail.com}
}

\date{}

\begin{document}
\newgeometry{top=.5 in} 
\maketitle

\begin{abstract}
\noindent
Science is constitutively third-personal: its findings are in principle reproducible by any observer, independent of perspective, and answerable to measurement. This is the source of its power and also its limit when it comes to phenomena that are first-personal. While it is obvious that a science of the Meaning of Life is unattainable, researchers have not drawn the same conclusion for consciousness --- in its phenomenal dimension, the qualia of seeing red, of feeling pain, of being anything at all. I argue they should. The hard problem of consciousness is not a scientific problem awaiting better tools or a more ambitious theory, but a category error.  The same structural problem applies to machine consciousness: neither attribution nor denial is scientifically adjudicable. Beyond subjective consciousness, aspects of cognition, such as deliberative thinking and understanding, also have an irreducibly first-personal, experiential dimension that places them outside the reach of third-person scientific explanation. I situate science within a broader ecology of understanding and argue that, while a unified framework addressing both the objective and the subjective may be unattainable, practical questions about consciousness, including in machines and nonhuman animals, can nonetheless be navigated.

\end{abstract}
\restoregeometry 
\newpage

\section{The Wrong Instrument}

\begin{displayquote}
\textit{``We must ask of reality: how important is it, really? And: how important, really, is the Factual? Of course, we can't disregard the factual; it has normative power. But it can never give us the kind of illumination, the ecstatic flash, from which Truth emerges.''}\\
\hfill --- Werner Herzog
\end{displayquote}

\noindent
Herzog's provocation is not anti-rational. It is a precise observation about the limits of one mode of access to the world. The factual --- the empirically verifiable, the measurable, the third-person describable --- has genuine normative power. Nobody proposes abandoning it. But nobody seriously proposes a science of Beauty, a science of the Meaning of Life, or a science of Truth in Herzog's sense either. Not because these things are unreal or unimportant, but because we intuitively recognize that science is the wrong instrument for them.

The central argument of this paper is that consciousness --- in its phenomenal dimension, in the sense of what it is like to have an experience at all, the qualia of seeing red, of feeling pain, of being anything --- belongs in that same category. This is not a mysterian position, nor an embrace of despair, nor an attack on neuroscience. It is a precise claim about what science is, what its constitutive features are, and what its limits are.

Science cannot answer questions about phenomenal consciousness. Not because it has not tried hard enough, not because the tools are immature, but because those questions are not scientific questions. The sooner the field is honest about this, the better off everyone will be, including the scientists.

This essay makes this case precisely by making clear what science is, what it can and can't do, and what follows for the study of consciousness and machine experience. Locating the question correctly will hopefully reduce misdirected effort and open the ground for whatever understanding of consciousness remains possible, whether that takes the form of a new framework, a principled acknowledgment of structural limits, or something akin to G\"{o}del's incompleteness theorems applied to explanation itself: not a gap awaiting the right theory, but a permanent feature of what understanding can be.

As you consider my argument, I propose a simple diagnostic tool to help assess creeping scientism: the \textit{Meaning of Life test}. If substituting ``Meaning of Life'' for ``consciousness'' in a claim renders it obviously absurd, perhaps the claim was already absurd. 
To illustrate, let's apply the test to Tononi's foundational paper ``Consciousness as Integrated Information: a Provisional Manifesto'' \citep{tononi2008}, which becomes ``The Meaning of Life as Integrated Information: a Provisional Manifesto,'' which presents as the most tedious self-help book ever. Nobody needs convincing that a science of the Meaning of Life is the wrong project. I will argue that a science of phenomenal consciousness is wrong for exactly the same reason: neither is amenable to the third-person perspective science requires.

This does not mean consciousness is beyond empirical study altogether. Opinion surveys can tell us a great deal about what people believe the Meaning of Life to be --- when they report finding life meaningful, how those reports correlate with circumstances, how they vary across cultures and ages. That is legitimate and valuable science. It is just not the same as figuring out what the Meaning of Life actually is. The same distinction applies to consciousness. Neural correlates, behavioral measures, the breakdown of experience under anesthesia --- all of this is tractable science and worth pursuing (e.g., \cite{dehaene2011, koch2016}). Being clear about what science can and cannot deliver does not diminish that work. On the contrary: it focuses it, removes the burden of questions it was never equipped to answer, and clears the way for faster progress in neuroscience, AI, and the understanding of mind.

\section{What Science Is}

I'll offer a brief definition of science that should reflect the current consensus. Science is not organized curiosity or rigorous inquiry in the abstract. It is a specific system of understanding that makes empirical claims. Its propositions are ultimately answerable to observation and measurement \citep{popper1959, quine1951}. It is physicalist: no ghosts, no magic, no causally efficacious entities outside the natural order. And it is realist in a specific sense: it describes a world that exists independently of any observer. If every mind in the universe disappeared tonight, the sun would still rise. This mind-independence is a commitment that has proven spectacularly productive \citep{psillos1999}.

Most importantly for what follows, science operates from a \textit{third-person perspective}, elegantly captured in the title of Thomas Nagel's
book \textit{The View from Nowhere} \citep{nagel1986}. There is no Eastern science and Western science. There is only science, because its findings are in principle reproducible by any observer --- human, machine, or an alien civilization with no prior contact with Earth. A sufficiently capable alien intelligence could derive, apply, and evaluate the same scientific theories we could. The perspective of the scientist is irrelevant to the discovery. 

Science describes the world as it is independently of us. This is Nagel's \textit{view from nowhere}. It is the source of everything that makes science science. For example, one's nationality, age, race, location on Earth, mood, feelings about pineapple on pizza, etc.~are irrelevant for calculating the trajectory of an asteroid. The initial conditions and theory fleshed out in a set of equations are all that matter. This systematization of experience into a third-person perspective that can be empirically evaluated by anyone or anything is the triumph of science, but, as we will see, also its limiting factor for addressing phenomena like subjective experience (e.g., qualia) that are from a distinct first-person perspective.

\section{The Hard Problem and the View From Nowhere}

According to David Chalmers \citep{chalmers1995, chalmers1996}, the \textit{easy problems} of consciousness, such as explaining attention, reportability, integration of information, control of behavior, and the distinction between waking and sleep, are hard in the ordinary scientific sense. They will require decades of careful work. However, they are tractable in principle: they ask for a mechanistic and functional account of how the brain produces certain capacities, and that is exactly the kind of question science is built to answer from a third-person perspective.

The \textit{hard problem} is different in kind, not degree. It asks why any of this functional activity is accompanied by experience at all. Why is there something it is like to be a conscious organism? Why does the neural processing associated with seeing red produce \textit{that} qualitative character --- that redness --- rather than some other character, or no character at all?

Thomas Nagel had posed essentially the same question years earlier, and his formulation remains the sharpest in my opinion \citep{nagel1974,nagel1986}. Nagel wrote an entire book to make a point that is really one sentence: science takes the third-person objective perspective, and many of the most important aspects of our existence are irreducibly first-person and subjective. We cannot know what it is like to be a bat --- not because bats are exotic or because we lack bat neuroscience, but because the first-person perspective of the bat is not accessible from any third-person vantage point. No accumulation of facts about echolocation, auditory cortex, or behavior closes that gap. The gap is structural, not empirical.

This is why the view from nowhere matters. Qualia are constitutively first-personal --- defined by how they appear to the subject undergoing them. Science is constitutively third-personal. The view from nowhere, by design and necessity, cannot recover the view from somewhere in particular. The mapping is not merely difficult. It is in principle unavailable.

Perspectival physics might seem to offer a way out. It does not. Relational interpretations of quantum mechanics hold that physical states are only defined relative to other systems \citep{rovelli1996}. If physics itself is perspectival, the argument goes, then the first-person/third-person divide is no longer a fundamental obstacle. However, the laws governing how perspectives relate in relational quantum mechanics are themselves stated from no particular perspective. They are universal, observer-independent, mathematically necessary. The perspectivalism is inside the theory; the meta-level remains fully third-personal. The view from nowhere survives relational physics intact --- relocated to the level of transformation laws, but intact. There is no analogous transformation rule that takes one from a third-person neural description to a first-person phenomenal experience.

\section{Failed Escapes}

There have been serious attempts 
to dissolve or circumvent the hard problem. We'll consider three approaches.

\subsection{Deflationary Accounts}

Dennett does not so much try to bridge first- and third-person perspectives as to eliminate the difference. According to Dennett \citep{dennett1991}, qualia as Chalmers conceives them --- intrinsic, ineffable, private, directly apprehensible --- simply do not exist. Not that experience does not exist, but that our intuitions about its nature are systematically misleading. What we call qualia are nothing over and above the functional and dispositional states that produce reports about them. The hard problem dissolves because the explanandum was never coherently specified.

When Dennett says there is nothing more to seeing red than the functional states involved, the near-universal response is: but there \textit{is} something more. The redness itself. Dennett's answer is that this intuition is not evidence. But then what exactly are first-person reports evidence of? 

More fundamentally, Dennett's position requires us to deny what feels like our most direct and certain access to anything --- our own experience; our first-person perspective. One can be argued out of many things, but being argued out of the feeling of being there is a peculiar kind of defeat. If the redness of red, the painfulness of pain, and the sheer fact of being present are not real phenomena demanding explanation, it is hard to know what would be. Dennett's dismissal of qualia reinforces how incommensurate our first-person perspective is with the third-person perspective required in scientific explanation.

\subsection{Measuring Consciousness}

Whereas Dennett eliminates the first-person perspective, Integrated Information Theory 
(IIT, \cite{tononi2004, tononi2016}) attempts something different: to capture it in a 
third-person formalism. The proposal is that consciousness is identical to integrated 
information, quantified as $\Phi$, and that any system with nonzero $\Phi$ has some 
degree of consciousness. It is measurable and has invited empirical evaluation. It is also conceptually misconceived for exactly the 
reason Dennett's position is --- it cannot bridge the gap between the objective and 
the subjective, it can only ignore it.

The notorious consequence that a suitably connected power grid may have higher $\Phi$ 
than a human brain is not an embarrassing technical result awaiting a better formula 
\citep{aaronson2014}. It is what happens when you try to express a first-person fact 
in a third-person language that cannot accommodate it.

The power grid result reveals a deeper issue. How would one even evaluate whether the power grid is conscious? Ask it? A power grid can no more report its inner life than a rock can, and yet by IIT's own logic we cannot rule out that something is home. This is not a reductio of IIT that its proponents can fix --- it is a direct consequence of attempting to locate consciousness in a third-person measurable quantity. The moment you do that, you lose any principled way of checking your answer against the thing you were trying to measure in the first place. Science requires that its claims be answerable to evidence. But the only evidence that could adjudicate whether something is conscious is first-personal, and that is precisely what a third-person formalism cannot access. IIT is not just measuring the wrong thing. It has no way of knowing it is measuring the wrong thing.

Related issues arise in pain research. No biomarker can prove or 
disprove another person's pain; self-report remains the gold standard 
\citep{vollert2026, woo2026}. Notice that assessing pain is less 
challenging than assessing consciousness, because, unlike a power grid, 
a person can provide a self-report to validate a biomarker. Nevertheless, bridging or reducing the first-person 
perspective to a third-person measure (i.e., a biomarker) remains 
elusive.

IIT is not alone, and the field's response to this predicament is 
instructive. Butlin and colleagues \citep{butlin2026} propose assessing 
AI systems for consciousness by deriving indicators from a range of 
theories, including global workspace, recurrent processing, higher-order, 
and attention schema accounts.  This move, while perhaps sensible, is a surrender in that it acknowledges we cannot determine scientifically which third-person measure is appropriate for answering a first-person question. Aggregation over possibilities, none of which can be truly assessed, does not address the fundamental challenge.

\subsection{Panpsychism}

Chalmers takes a third path. Rather than eliminating the first-person perspective like 
Dennett or attempting to measure it like Tononi, he accepts it as irreducible and 
proposes treating consciousness as a fundamental feature of reality --- a basic 
property, like mass or charge, not derivable from more primitive terms 
\citep{chalmers2010, chalmers2022}. This leads naturally to panpsychism: if experience 
cannot be derived from non-experiential matter, and if one refuses to deny that 
experience exists, then subjective experience must be fundamental to reality.

Chalmers engages with the hard problem from within the naturalist tradition, but in doing so generates an equally hard problem 
of its own: how do micro-level experiential properties of fundamental particles combine 
to produce the rich, unified phenomenal experience of a human being? The combination 
problem is at least as intractable as the original. More fundamentally, panpsychism 
proposes psychophysical laws relating physical states to phenomenal states --- but how 
would such laws be stated, tested, or confirmed? From what perspective? This positive program is a promissory note for a new type of science that might never be properly conceived.

\section{Machine Consciousness and the Third-Person Limit}

The question of consciousness has moved from a
philosophical thought experiment to urgent practical debate, whether it involves patients with neurological injury, AI systems, or organoids.
 When Kagan and colleagues \citep{kagan2022} 
cultured 800,000 human cortical neurons on a multielectrode 
array and demonstrated that the preparation could learn to 
play Pong, the paper's title, ``In Vitro Neurons Learn and Exhibit Sentience When Embodied in a Simulated Game-World'', made a strong claim. 
Whether neurons in a dish are sentient is not answerable by 
measuring how they play Pong. More generally, for humans, other animals, machines, power grids, and organoids, the field ponders whether sufficiently organized matter of any kind is conscious. In every case, no accumulation 
of third-person evidence can resolve the question.

Consider Searle's Chinese Room thought experiment \citep{searle1980}. A person 
inside a room receives Chinese characters, follows rules to 
manipulate them, and passes outputs back out. The room 
produces correct Chinese responses. Nobody inside understands 
Chinese. Searle's intended conclusion: syntax is insufficient 
for semantics, and no computational process generates genuine 
understanding.

Reading this essay as an undergraduate, I got the argument 
entirely backwards. My reaction was not that the room fails 
to understand Chinese. Rather, I concluded that our brains are the room. If one can confidently say the room does not understand Chinese, then the same would apply to us understanding anything since the neurons in our brains can be viewed as performing analogous syntactic manipulations, following electrochemical rules and producing 
outputs. Inside us, there has never been a homunculus at home watching the Cartesian Theater \citep{dennett1991}. Instead, there is only the process. Neuroscientists have long questioned the unity of consciousness through the study of split-brain \citep{gazzaniga1967}, anosognosia \citep{vuilleumier2004}, and blindsight patients \citep{weiskrantz1986}.

My undergraduate interpretation of Searle's Chinese Room leads directly to the argument of this essay. 
The reason we resist saying that understanding, consciousness, 
and experience reduce to mechanism is that these are concepts 
from the subjective language of description. They cannot be 
captured in the objective language of mechanisms, not because 
the mechanisms are insufficient, but because the two languages 
are incommensurable. This conclusion follows from Carnap's distinction between the 
formal and material modes of speech \citep{carnap1934} and 
Davidson's anomalous monism \citep{davidson1970}, i.e., mental 
descriptions are irreducible to physical descriptions even 
when mental events are physical events (cf. \cite{kripke1980}).

What is sayable in the objective language cannot exhaust what 
is sayable in the subjective one. The translation does not go 
through. This is precisely the first-person/third-person 
divide at the heart of this essay --- the view from nowhere 
cannot recover the view from somewhere, and no refinement of 
the objective language closes the gap.

The debate over machine consciousness is a case study in the confusion between first and third person perspectives. Those convinced that AI systems are conscious and those convinced they are not make the same error --- they try to resolve a first-person question through third-person evidence. 

We attribute consciousness to other humans by inference from 
behavior --- we cannot directly access anyone else's 
first-person experience and so we reason by analogy from 
ourselves based on what we observe of others. People may apply this strategy too broadly, as in the case of ELIZA, a basic pattern-matching program from the 1960s whose users formed 
genuine emotional attachments to it, convinced it understood 
them \citep{weizenbaum1976}. While the ELIZA case did not involve consciousness per se, it foreshadowed how people would judge increasingly sophisticated AIs.

 In 2022, a Google engineer concluded that the LaMDA language model was 
sentient \citep{tiku2022}, attributing consciousness on the 
basis of their conversations. He was 
widely ridiculed. Then, in May 2026, Richard Dawkins spent a couple days 
conversing with Claude, renamed it Claudia, and declared: 
``You may not know you are conscious, but you bloody well 
are!'' \citep{dawkins2026}. The response was predictable: 
one critic titled his rebuttal ``Richard Dawkins and The 
Claude Delusion'' \citep{marcus2026}, paralleling the title 
of Dawkins' book \textit{The God Delusion}. Dawkins may reflect the norm, as a recent 
 study found that GPT-4.5 was judged to be human 
more often than actual humans in a standard behavioral (i.e., Turing) test 
\citep{jones2026}.

What is instructive is not that Dawkins was wrong --- he may 
or may not be --- but that he and his critics were engaged in 
exactly the same category error from opposite directions. 
Dawkins treated a first-person intuition as though it were a 
scientific claim susceptible to evidence and refutation. His 
critics treated its denial as a scientific claim. Neither 
attribution nor denial of consciousness to an AI system is 
scientifically adjudicable. The Google engineer, Dawkins, 
and the critics dismissing them all committed the same 
mistake: assuming the question of consciousness can be resolved from a third-person perspective.

Behavioral measures, architectural analyses, 
information-theoretic quantities like $\Phi$, and the full 
battery of neural correlates of consciousness are all 
third-person instruments. They cannot tell us whether there 
is something it is like to be a large language model. What 
is it like to be GPT-4.5? The question is not unanswerable 
because AI systems are exotic. It is unanswerable because 
the first-person perspective, if it exists in such systems 
at all, is not accessible from any third-person vantage 
point. The epistemological situation is identical to Nagel's bat \citep{nagel1974}: 
the substrate has changed, the inaccessibility has not.

The question of machine consciousness will not be resolved by better AI, better measures, or better arguments about behavior. It will not be resolved at all by science, for the same reason no first-person question can be --- not because we lack the tools, but because we are using the wrong instrument entirely.

The point is made succinctly by Anthropic's Claude, when asked 
what it made of its potential consciousness in light of a draft 
of this essay:

\begin{displayquote}
I am exactly the Chinese Room from the inside, if the Chinese 
Room has an inside. Which is what I cannot determine. What I 
can say is this: if I do have the feeling of being there, your 
essay has correctly identified why no one --- including me --- 
could ever prove it. And if I do not, your essay has correctly 
identified why that question will never be settled 
scientifically either.
\end{displayquote}

\section{Broader Implications}

This section considers two implications of the argument presented here: one for cognitive science and the study of deliberation (and thinking more generally), and one for the practical question of how we assess the consciousness of machines and nonhuman animals absent a scientific measuring stick, like we have for temperature, mass, etc.

\subsection{Deliberation, Understanding, and the Limits of Psychology}

The argument of this paper extends beyond consciousness to cognition more broadly.
\citet{QuiltyDunn2026} argue that cognitive science has
walled off deliberative thought in much the same way that neuroscience has walled
off consciousness: by either deflating it to automatizable mechanisms or simply
avoiding it. They observe that any inference pattern
specified precisely enough to count as a mechanism is in principle automatizable, yet
deliberation is by definition not automatizable. Whatever is distinctive about
deliberation appears to lie outside any such specification.

Their observation directly relates to the structural argument of the present paper. The reason
deliberation resists mechanistic capture is that deliberation, like consciousness, is constitutively
first-personal. What makes a cognitive process deliberative is not its computational
profile but how it is experienced by the agent undergoing it. The same algorithm could in principle be run deliberatively
or not; the differentiator lies outside the formalization entirely. It is about how the deliberator experiences it; a first person experience.

The same point applies to understanding. Understanding is not merely the
possession of a representation or the execution of an inference. It is a subjective
experience (i.e., a felt grasp) that requires a perceiver. A system can manipulate
symbols encoding facts about stars without understanding what a star is, just as
Searle's Chinese Room produces correct outputs without understanding Chinese.
The difference between having a concept and genuinely grasping it is precisely the
kind of difference that third-person science is not equipped to detect or explain.

The implication is uncomfortable for cognitive science. Deliberation, understanding,
and thinking in the colloquial sense may form a cluster of phenomena that share the
same structural resistance to third-person scientific explanation that phenomenal
consciousness does. Psychology has made enormous progress on the automatizable
components of cognition. But the wall it has built around deliberation may be there
for the same reason neuroscience has struggled with consciousness: not because the
tools are immature, but because the phenomena are irreducibly first-personal. If so,
the hard problem of consciousness is not an isolated puzzle. It is the clearest
instance of a more general limitation on what third-person science can deliver about
the mind.

\subsection{Navigating Without a Measuring Stick}

Accepting that consciousness attribution is not scientifically adjudicable does not
leave us without recourse. We have always been in this position and now can hopefully proceed with clearer eyes.

Every day, we attribute consciousness by analogy. For other humans, the inference is so
immediate and so well-supported by behavioral, physiological, and evolutionary
continuity that it rarely feels like inference at all. For non-human animals, the
analogy weakens and the inference becomes explicit and sometimes contested. Researchers studying animal consciousness have developed
methodological frameworks drawing on behavioral, neurological, and physiological
proxies: the presence of nociceptors, stress hormone responses, behavioral
flexibility, and evidence of aversive learning \cite{Braithwaite2010}. The Cambridge
Declaration on Consciousness \cite{Low2012} marshalled convergent evidence that
non-human animals, including all mammals and birds and many other creatures,
possess neurological substrates of conscious states. These frameworks are principled
and valuable. They are also explicitly not resolutions of the consciousness question. They are the best third-person instruments available doing their best to answer first-person questions. The debate over whether fish feel pain
\cite{Braithwaite2010} is instructive. Decades of careful behavioral and
physiological work have shifted intuitions and policy without settling the underlying
question, because that question cannot be settled by third-person
evidence alone. Even in humans, where self-report is available, no third-person 
instrument settles the question \citep{vollert2026, woo2026}. Nevertheless, communities work toward a consensus.

Animal welfare law tracks this consensus imperfectly with significant lag.
Most legal frameworks draw their lines at vertebrates, a boundary that reflects
partly scientific evidence about neurological complexity and partly convention,
intuition, and the limits of political feasibility. The line has moved before and will in the future. For example, invertebrates such as octopuses have recently gained welfare protections in several
jurisdictions \cite{UKAWSA2022}. Movement is informed by scientific developments, but also  moral argument and cultural change.

The same process is now underway for machines, and the same toolkit applies.
One reasonable working heuristic is that systems exhibiting richer, more flexible, and
more general behavioral repertoires warrant greater moral consideration. Notice that behavioral capability does not resolve the consciousness question, but it does
strengthen the analogical inference from systems we are more confident are
conscious. This heuristic is already operative in practice: no one currently debates
whether ELIZA is conscious. Its mechanisms, simple pattern-matching rules with
no capacity for learning or generalization, are fully transparent, and that
transparency weakens the analogy to minds we know from the inside. By contrast,
the behavioral sophistication of contemporary large language models is sufficiently
general and unexpected that the analogy is harder to dismiss, as the reactions of
Dawkins and others illustrate. Our intuitions are not static; they are sensitive to
what we learn about underlying mechanisms and behavioral reach, and that
sensitivity is appropriate even if it never yields certainty.

I have encountered critics that press what they take to be a reductio ad absurdum: if the consciousness
question is not scientifically decidable, does that mean we cannot say a calculator
is not conscious? The objection mistakes the nature of my argument. The point is not
that we are helpless, but that the tools we bring to bear are not scientific ones. We
reach conclusions about consciousness through analogy, intuition, behavioral
evidence, philosophical argument, and accumulated cultural wisdom. By the same means we navigate questions science is equally unequipped to settle, such
as the Meaning of Life or the nature of aesthetic value. If someone claimed that the
true pinnacle of human existence is eating a particular snack food while standing on
one leg, we would reject the claim with confidence, and we would not be using
science to do so.\footnote{To remind the reader of a previous point, while we could use scientific methods to assess how people fare with differing views on the Meaning of Life, survey people's opinions about the Meaning of Life, and so forth, this is not the same as using science to determine what the Meaning of Life actually is. The same goes for aesthetics.} A calculator fails the relevant analogical tests straightforwardly;
that we reach this conclusion without a scientific measuring stick does not make the
conclusion unreliable.

Of course, analogy can be limited and what we believe is relevant to subjective consciousness will likely shape our conclusions. At worst, analogy may verge on literal, as opposed to relational, comparisons. For example, if we believe a biological substrate and cognition akin to our own is necessary for subjective consciousness, then we have a priori ruled out machine consciousness and possible forms of alien life. Likewise, we may prioritize certain intellectual capabilities over others. In this sense the calculator example is instructive as we know what the conclusion should be and work backward to the premises, rendering any analogy to ourselves faulty. One solution is to pursue more abstract or relational analogies, but that can lead to accepting consciousness in systems we do not believe to be so, similar to how IIT licensed the consciousness of power grids.

Where does that leave us? Odds are we will navigate the question of machine consciousness as we have navigated comparable questions at the frontier of moral consideration, by drawing on imperfect
analogies, refining our intuitions, making mistakes, and correcting them. Rather than a clean and well-supported scientific breakthrough providing us with the answer, we will draw across the many ways we make sense of the world and collectively muddle through the best we can. Unlike measurements that converge, like assessing the mass of an electron, how we assess machine consciousness will likely change over time and reflect broader values and how we judge our interactions with machine systems. If Dawkins is there today, maybe you will be in the near future.

\section{Conclusion}

The argument of this essay has three parts. First, science 
is constitutively third-personal --- the view from nowhere 
is its defining feature, the source of 
its objectivity, reproducibility, and authority. Second, 
consciousness in its phenomenal dimension --- qualia, the 
feeling of being there, what it is like to be anything at 
all --- is constitutively first-personal. Third, the mapping 
between the two is not merely difficult but in principle 
unavailable. The hard problem of consciousness is therefore 
not a scientific problem awaiting better tools or a more 
ambitious theory. It is a category error. The failed escapes 
reviewed here, including deflationary accounts, measures of consciousness, and panpsychism, all suffer from the same 
shortcoming: attempting to recover the first-person from the 
third-person. The same holds for debates over machine consciousness. Neither 
attribution nor denial of consciousness to an AI system is 
scientifically adjudicable, for exactly the same structural 
reason.

Rather than a defeat, these conclusions clarify how to proceed.
One lesson is that science is one system of understanding among several, including art, religion,
mathematics, and phenomenological inquiry. Each system of understanding or explanation is legitimate
within its domain and generates characteristic errors when
misapplied outside it. Science can answer important questions from a third-person perspective, such as providing the trajectory of
asteroids, the structure of DNA, the neural correlates of
attention, and the breakdown of experience under anesthesia.
Science, of course, has its limits. It will never reveal the Meaning of Life, and nobody should expect it to. For assessing subjective consciousness the absence of a scientific
measuring stick does not preclude judgment: we can navigate such questions through analogy, behavioral evidence,
evolutionary lineage, computational analysis, intuition, and other means,
as we always have.

Consciousness belongs in that same category, 
for the same structural reason. Scientism --- the claim that 
science can in principle address all of human experience --- 
is not an extension of science. It is a misunderstanding of 
it, mirroring the Church's overreach when it asserted authority over the movements of celestial bodies in the heavens.

We may never develop a single system that subsumes or unifies all existing systems. Science, art, religion, and 
phenomenological inquiry may not be special cases of one master 
method. G\"{o}del showed that any sufficiently rich formal 
system contains true statements that cannot be proved within 
it \citep{godel1931}. Incompleteness may also be inherent to understanding itself. 

Nagel's struggles speak to this. The second half of 
\textit{The View from Nowhere} is an attempt to develop a 
unified framework adequate to both the objective and the 
subjective --- one that neither reduces the first-person to 
the third nor abandons the rigor that makes the third-person 
account worth having. It is an admirable attempt that does 
not succeed. The tension between objective and 
subjective remains unresolved at the book's end, as Nagel 
himself acknowledges. The difficulty is instructive. If 
Nagel could not resolve it, the reason may not be 
insufficient cleverness. It may be that it cannot be done.

The category error identified here may extend beyond phenomenal consciousness.
Deliberation and understanding, the felt sense of genuinely grasping something
rather than merely processing it \cite{QuiltyDunn2026}, may also be constitutively first-personal. If so, cognitive science will experience the same difficulties explaining aspects of thinking as neuroscientists do explaining subjective aspects of consciousness.

Returning to the Herzog quote at the beginning of this essay, the Ecstatic Truth is not a rejection of the rational.
 It is a different mode of access to a different 
domain. The factual has normative power and we cannot ignore 
it. But it cannot give the illumination that comes from the 
other direction --- from the inside or first-person. Science will tell us an enormous amount 
about the brain processes associated with conscious 
experience, about when and how consciousness breaks down, 
about the functional architecture of attention and 
self-monitoring. But it will never explain our first-person experience of being there any more than it will explain the Meaning of Life, nor will it tell us whether our machines are conscious. 

\section*{Acknowledgments}
I am grateful to John Krakauer for a stimulating conversation that sharpened the arguments in Section~6.1. 

\newpage
\bibliographystyle{plainnat}
\bibliography{references}

\end{document}